\documentclass[11pt]{article}
\usepackage{moriond}
\usepackage{graphicx}
\usepackage{color}
\usepackage{latexsym}
\usepackage{nicefrac}

\def\Journal#1#2#3#4{{#1} {\bf #2}, #3 (#4)}

\newcommand{\NPB}{{\em Nucl. Phys.} B}
\newcommand{\PLB}{{\em Phys. Lett.}  B}

\newcommand{\PRD}{{\em Phys. Rev.} D}

\newcommand{\CPC}{{\em Comp. Phys. Comm.}}
\newcommand{\LEPII}{{\sffamily LEP~II}}

\newcommand{\slant}[1]{#1\hspace{-1.7ex}{\red/}}
\newcommand{\LEP}{{\sf LEP}}

\newcommand{\ALEPH}{{\sf ALEPH}}
\newcommand{\LD}{{\sf L3}}
\newcommand{\OPAL}{{\sf OPAL}}
\newcommand{\DELPHI}{{\sf DELPHI}}
\newcommand{\SM}{Standard Model}
\newcommand{\MC}{Monte Carlo}
\newcommand{\CP}{{\sf CP}}
\newcommand{\CL}{{\sf CL}}
\newcommand{\SU}{SU(2)$_{\rm L}\times$U(1)$_{\rm Y}$}
\newcommand{\Uem}{U(1)$_{\rm em}$}

\renewcommand{\gg}{\ensuremath{g_1^\gamma}}
\newcommand{\gz}{\ensuremath{g_1^{{\rm Z}}}}
\newcommand{\kg}{\ensuremath{\kappa_\gamma}}
\newcommand{\kz}{\ensuremath{\kappa_{{\rm Z}}}}
\renewcommand{\lg}{\ensuremath{\lambda_\gamma}}
\newcommand{\lz}{\ensuremath{\lambda_{{\rm Z}}}}
\newcommand{\tws}{\ensuremath{\tan^2\theta_{\rm W}}}

\newcommand{\Oa}{\ensuremath{\mathcal{O}(\alpha)}}

\definecolor{orange}{rgb}{0,0,0}
\newcommand{\orange}{\color{black}}
\newcommand{\red}{\color{black}}
\newcommand{\green}{\color{black}}
\newcommand{\blue}{\color{black}}
\newcommand{\cyan}{\color{black}}
\newcommand{\magenta}{\color{black}}

\newcommand{\ee}{\ensuremath{\rm e^+e^-}}
\newcommand{\eeto}{\ensuremath{\rm e^+e^- \!\! \rightarrow }}

\newcommand{\eeZgto}{\ensuremath{\rm e^+e^- \rightarrow Z/\gamma^*\rightarrow \;}}
\newcommand{\eeWWto}{\ensuremath{\rm e^+e^- \rightarrow W^+W^-\rightarrow \;}}

\newcommand{\qqgamma}{\ensuremath{\rm q\bar q \gamma}}
\newcommand{\nngamma}{\ensuremath{\nu \bar\nu \gamma}}

\newcommand{\qqqqww}{\ensuremath{\rm q_1 \bar{q}_2 q_3 \bar{q}_4}}

\begin{document}
\vspace*{4cm}
\title{MEASUREMENT OF BOSON SELF COUPLINGS AT LEP AND SEARCH FOR
ANOMALIES}

\author{Martin Weber}

\address{III.\ Physikalisches Institut, Physikzentrum, RWTH Aachen,
D-52056 Aachen, Germany}

\maketitle

\abstracts{With center of mass energies up to 209~GeV of \LEPII,
massive W and Z bosons can be produced via $\ee$ collisions in pairs
and jointly with photons. This allows to study boson-boson
couplings. Since the W and Z bosons are unstable and decay into
fermions, two- and four-fermion final states, accompanied possibly by
photons, play an important role for these measurements. The couplings
of the W to other bosons have been measured to be $\gz =
0.990^{+0.023}_{-0.024}$, $\kg = 0.896^{+0.058}_{-0.056}$, and $\lg =
-0.023^{+0.025}_{-0.023}$. They are in agreement with the \SM{}
expectation of $\gz = 1$, $\kg = 1$, and $\lg = 0$. No sign for
couplings of three neutral bosons, parametrized by
$f_{4,5}^{\gamma,{\rm Z}}$ and $h_{1,2,3,4}^{\gamma,{\rm Z}}$, and for
anomalous couplings of four gauge bosons, parametrized by $a_0, a_n$
and $a_c$ has been found.}

\section{Couplings of the W to other bosons}

\begin{figure}[htbp]
\parbox{0.3\textwidth}{
\centerline{\includegraphics*[width=0.25\textwidth]{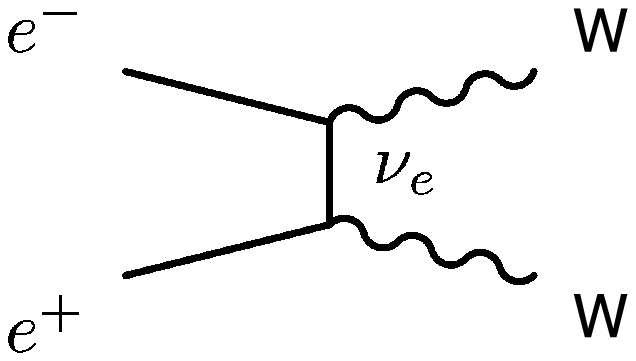}}
\centerline{\includegraphics*[width=0.25\textwidth]{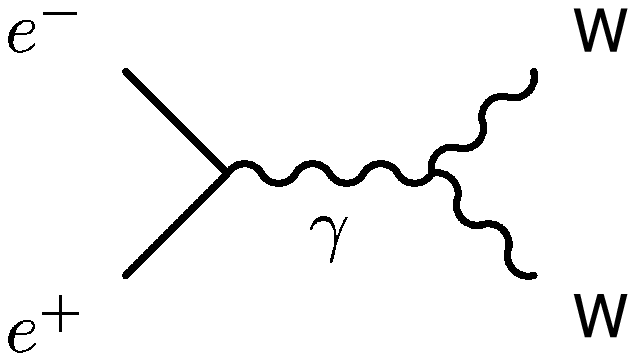}}
\centerline{\includegraphics*[width=0.25\textwidth]{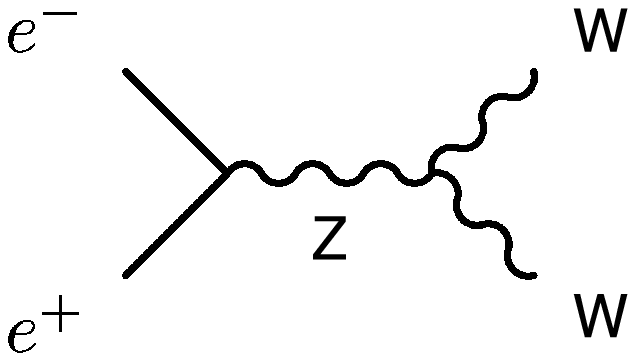}}
}\hfill
\parbox{0.65\textwidth}{
\includegraphics*[width=0.5\textwidth]{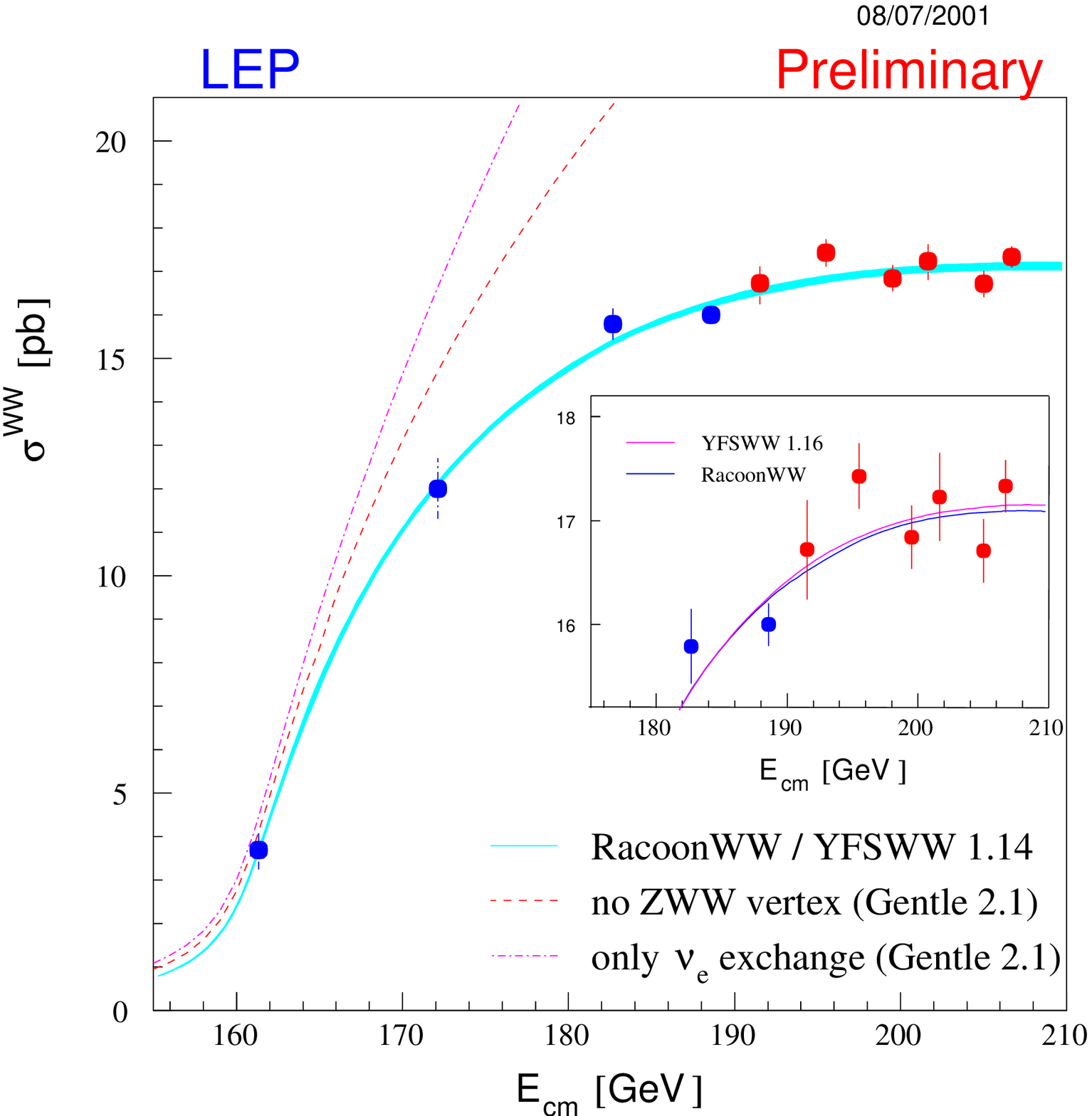}
\label{fig:ww-xsection}
}
\caption{Feynman graphs (left) and measured cross-section (right) for
the pair production of W bosons.}
\label{fig:ww-couplings}
\end{figure}

The \SU\ symmetry of the \SM\ predicts the pair production of W bosons
through Abelian and non-Abelian graphs. On the left side of
Fig.~\ref{fig:ww-couplings}, the three \SM\ feynman diagrams for W
pair production are shown.

As has been measured by the \LEP{} experiments, all three diagrams are
needed to describe the data. This can be seen on the right of
Fig.~\ref{fig:ww-couplings}. Using only the single Abelian graph (the
neutrino exchange) or neglecting the non-Abelian Z exchange graph,
data and theory disagree. But still the contribution of the graphs
could differ from the \SM\ prediction, and therefore a more
sophisticated method is performed to analyze the non-Abelian gauge
sector.

To study possible other contributions, the Lagrangian for the VWW
vertex (V=Z,$\gamma$) can be written in the most general Lorentz
invariant form~\cite{hagiwara}
\begin{eqnarray*}
i \mathcal{L}^{WWV}/g_{WWV} &=&
{\red g_1^V }
\left(
W_{\mu\nu}^\dagger W^\mu V^\nu - W_\mu^\dagger V_\nu W^{\mu\nu}
\right)
+
{\green \kappa_V} W_\mu^\dagger W_\nu V^{\mu\nu} \\
&+&
\frac{\blue \lambda_V}{m_W^2} W^\dagger_{\mu\nu} W^{\nu}_{\rho} V^{\rho\mu}
+ \slant{C} + \slant{P} + \slant{C}\slant{P}, 
\end{eqnarray*}
where {\sf C}, {\sf P} and {\sf CP}-violating terms are not shown and
assumed to vanish in the following discussion. To further reduce the
parameter set from six to three free couplings, firstly \Uem\ gauge
invariance is required, fixing the charge of the W boson to $q_{\rm W}
= \pm 1$, which is equivalent to $\gg = 1$. Secondly, the requirement
of \SU\ symmetry of the Lagrangian leads to the two constraints $\kz =
\gz - (\kg - 1) \tws$ and $\lz = \lg$. The three parameters left are
$\gz$, $\kg$ and $\lg$. In the \SM, their values are predicted to be
$\gz = 1$, $\kg = 1$ and $\lg = 0$. Often one finds in the literature
also the differences to the \SM\ expectations: $\Delta\gz = \gz - 1$
and $\Delta\kg = \kg - 1$.

\begin{figure}[b!]
\hspace{0.15\textwidth}
\parbox{0.25\textwidth}{
\includegraphics*[width=0.25\textwidth]{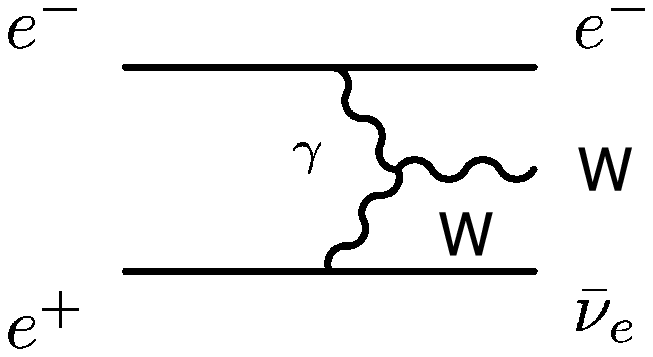}
\centerline{single W production}
}
\hspace{0.15\textwidth}
\parbox{0.25\textwidth}{
\includegraphics*[width=0.25\textwidth]{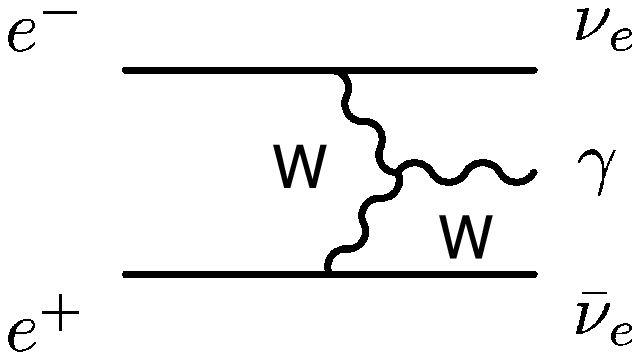}
\centerline{single $\gamma$ production}
}
\caption{Other processes that are used in the determination of the VWW
couplings.}
\label{fig:sw-and-sg}
\end{figure}

The couplings are not only accessible in W pair production, but also
in single W and single photon production, which also involve the
$\gamma$WW vertex, as can be seen from Fig.~\ref{fig:sw-and-sg}. The W
pair production is most sensitive to the couplings \gz\ and \lg, and
its sensitivity to \kg\ is comparable to the single W production,
which in turn is most sensitive to \kg. From all processes, the single
photon production is least sensitive.

Deviations from the couplings as they are predicted by the \SM{} would
lead to changes of the total cross section, of the production and
decay angles and of the average polarization of the bosons.

\begin{figure}
\includegraphics*[width=0.3\textwidth]{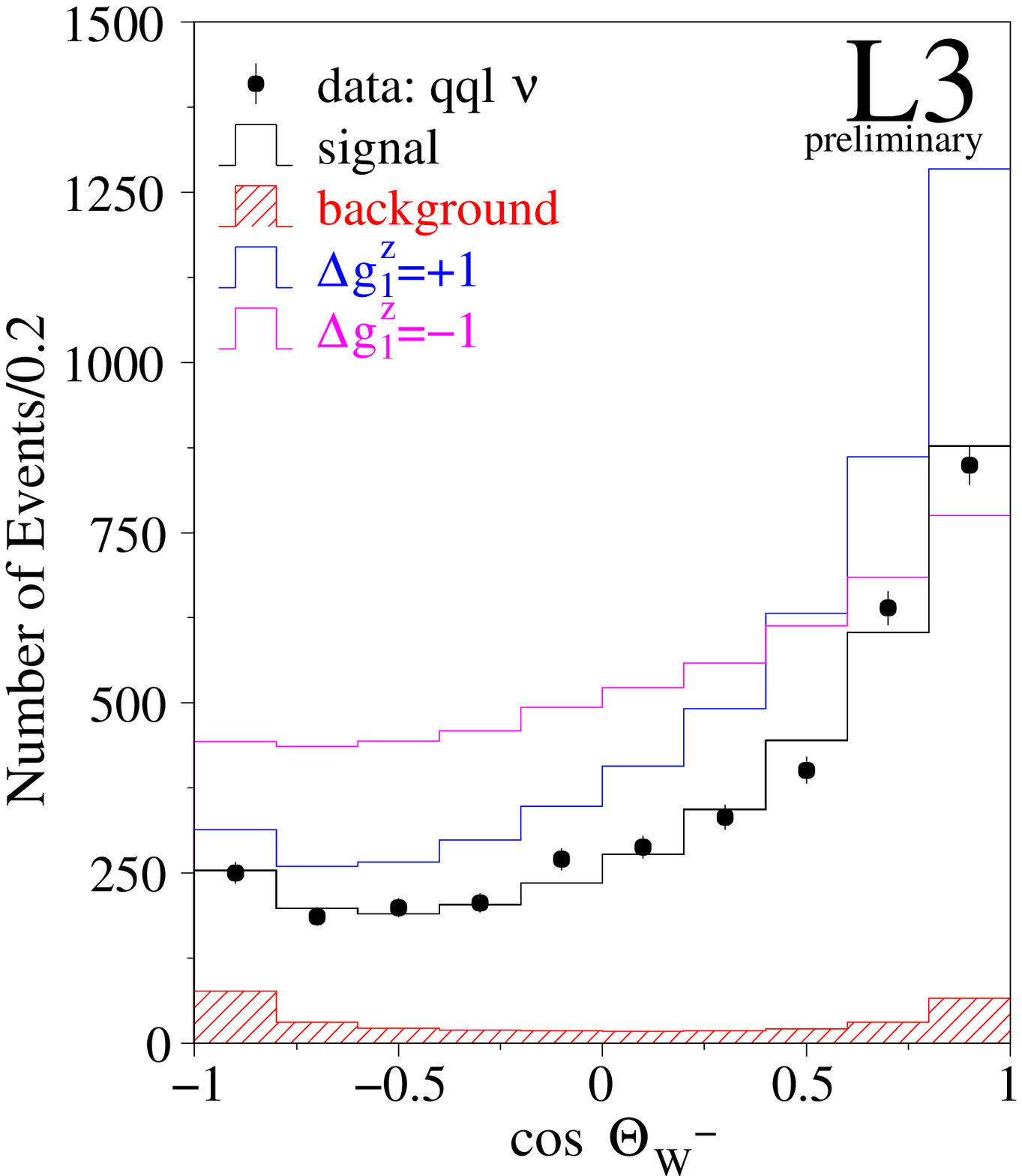}\hfill
\includegraphics*[width=0.3\textwidth]{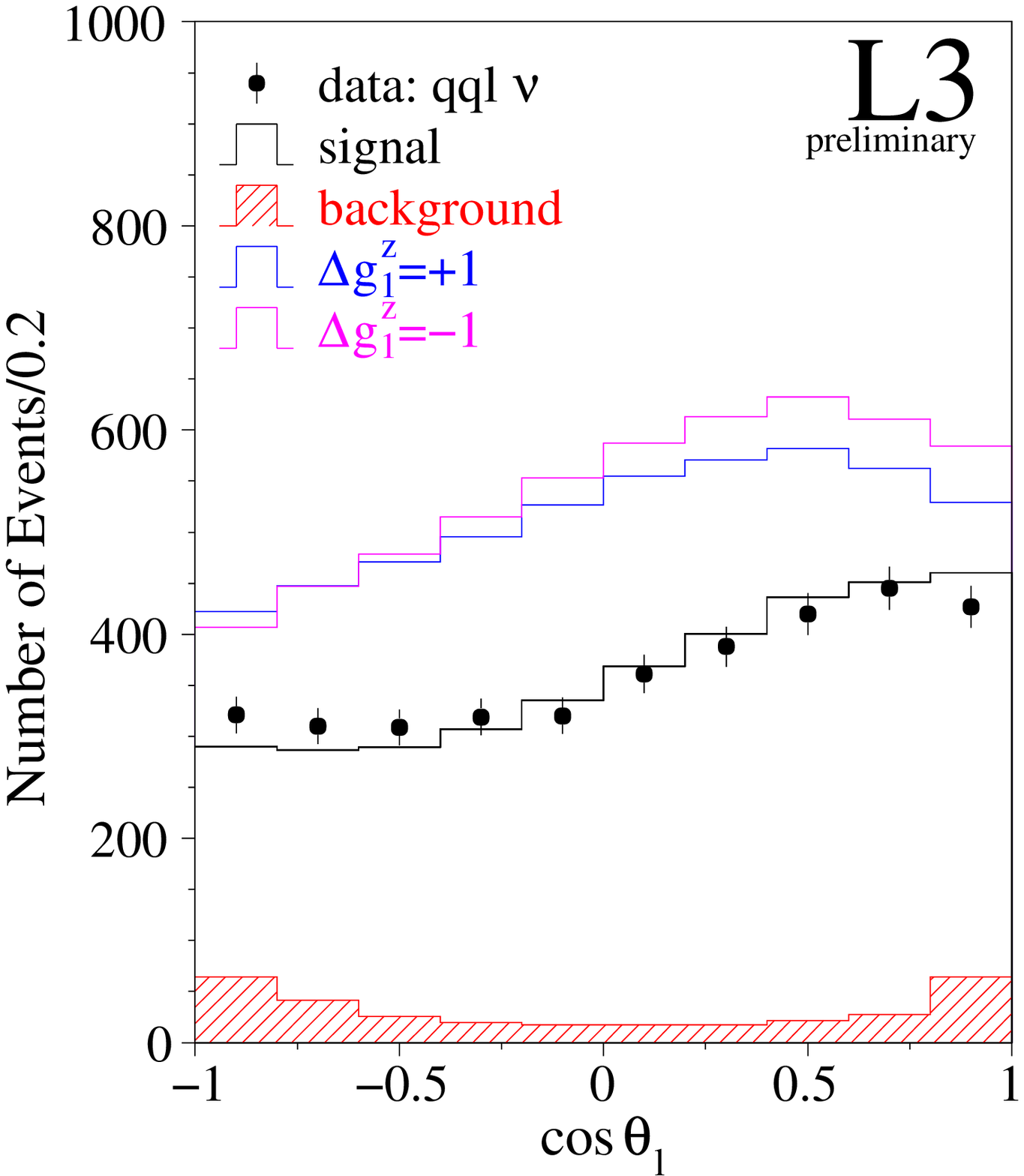}\hfill
\includegraphics*[width=0.3\textwidth]{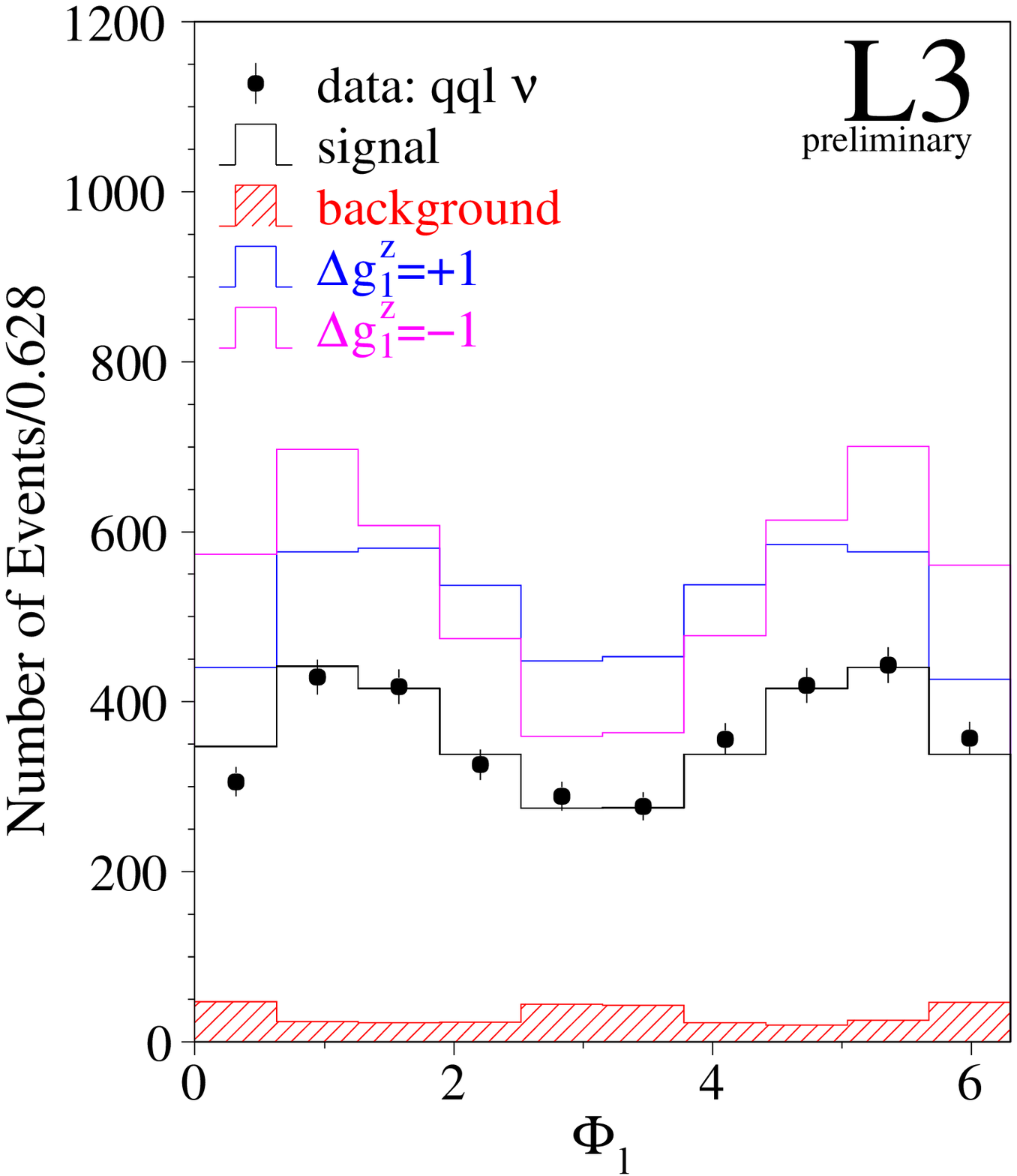}
\caption{Distribution of the production angle of the W boson and of
the decay angles of the lepton in the rest frame of the decaying W in
the semileptonic case. The angles for the quarks are not shown.}
\label{fig:ww-angles}
\end{figure}

In the W pair production process, all information about production and
decay is contained in five variables: The production angle
$\theta_{{\rm W}^-}$ of the ${\rm W}^-$, the polar and azimuthal
angles $\theta,\phi$ of the decay products in the rest frame of the
decaying ${\rm W}^-$ and ${\rm W}^+$ relative to the W flight
direction. If a W decays into a charged lepton and a neutrino,
$\theta$ and $\phi$ are taken from the charged lepton, and if a W
decays into two quarks, the angles are symmetrized to compensate the
missing charge determination. The distributions of $\cos\theta_l$,
$\phi_l$ and $\cos\theta_{{\rm W}^-}$ in the semileptonic case are
shown in Fig.~\ref{fig:ww-angles} as they have been measured by the
\LD\ experiment and together with the expectations for
$\gz=0,1,2$. From the shape of these distributions and the total rate,
constraints on the value of the couplings are derived.

\begin{figure}[b]
\includegraphics*[width=0.18\textwidth]{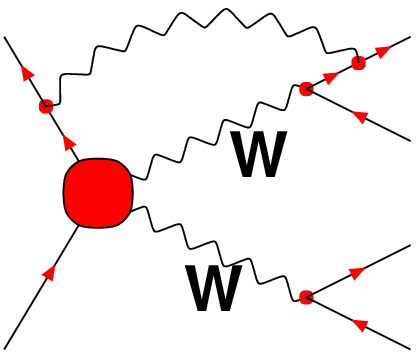}\hfill
\includegraphics*[width=0.18\textwidth]{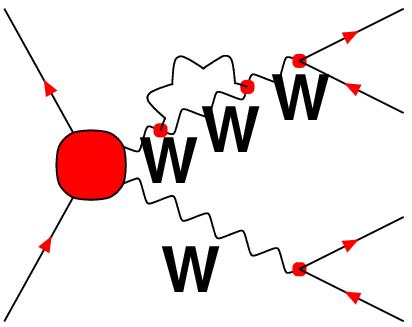}\hfill
\includegraphics*[width=0.18\textwidth]{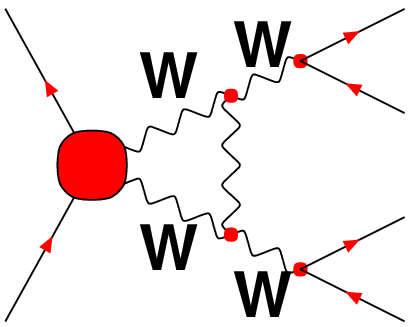}\hfill
\includegraphics*[width=0.18\textwidth]{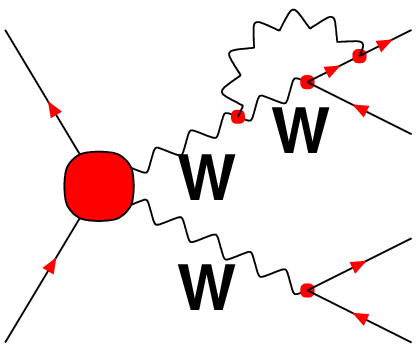}\hfill
\includegraphics*[width=0.18\textwidth]{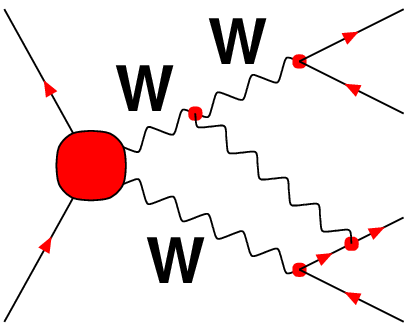}
\caption{Some example diagrams for \Oa corrections.}
\label{fig:oalpha}
\end{figure}

The shape of the $\cos\theta_{{\rm W}^-}$ distribution shows stronger
distortions than the shape of the $\cos\theta_l$ and $\phi_l$
distributions, if the couplings are changed. Therefore, a reliable
calculation of these distributions is necessary. Until recently, the
theory error was 2\% on the rate and larger for the differential
distributions like $\cos\theta_{{\rm W}^-}$, thus deteriorating the
measurement of the gauge couplings. By using the predictions from the
newly developed \MC\ generators {\sf YFSWW3}~\cite{yfsww} and {\sf
RacoonWW}~\cite{racoon}, a theory error of .5\% on \lg\ has been
achieved\cite{bruneliere}. The two generators take into account
\Oa{}-corrections, i.\,e.\ diagrams with internal and external photon
lines, in the Leading Pole Approximation~({\sf LPA}) and the Double
Pole Approximation {\sf DPA}, respectively. Some example diagrams of
these corrections are shown in Fig.~\ref{fig:oalpha}.

\begin{figure}
\includegraphics*[width=0.99\textwidth]{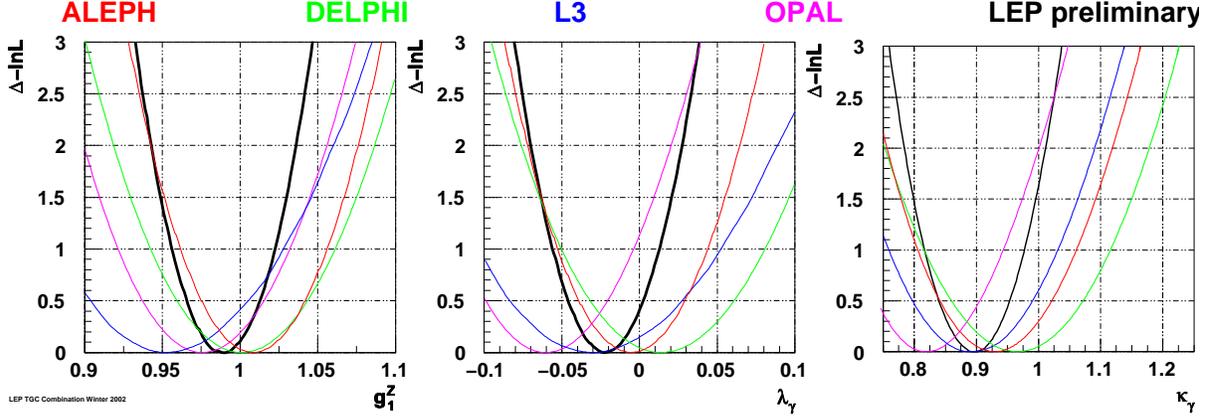}
\caption{Result of the triple gauge coupling fit.}
\label{fig:couplings-results}
\end{figure}

By using the predictions from {\sf YSFWW3}, measurements of the
couplings are performed by each experiment, and combined with a
log-likelihood method~\cite{alcaraz,lep-gc}. The likelihood curves of
the combined fit are shown in Fig.~\ref{fig:couplings-results}. The
measurement of \kg\ agrees within two standard deviations with the
\SM, and both \lg\ and \gz\ agree within one standard deviation with
the \SM. The fitted values with the errors corresponding to $\Delta L
= 0.5$ are\cite{lep-gc}:
\begin{displaymath}
\gz = 0.990^{+0.023}_{-0.024} \hspace{1cm} 
\kg = 0.896^{+0.058}_{-0.056} \hspace{1cm}
\lg = -0.023^{+0.025}_{-0.023}
\end{displaymath}

For this combination, both the \LD\ and \OPAL\ experiments did not
submit the \eeWWto \qqqqww\ channel. By adding these channels, the
statistical accuracy of the measurement will improve. As far as
systematic uncertainties are concerned, the \Oa{} corrections are the
largest correlated ones ($\pm 0.039$ on \kg, $\pm 0.015$ on \lg\ and
$\pm 0.015$ on \gz). For the result shown above, they have been set to
the full difference between the \MC{} prediction with and without
\Oa{} corrections. More refined numbers will be used in the future,
but are not available yet. Also, updates on the fits of higher
dimensionality relating two or three couplings are planned.

\section{Couplings of three neutral bosons}

\begin{figure}[b]
\hspace{0.15\textwidth}
\includegraphics*[height=0.15\textwidth]{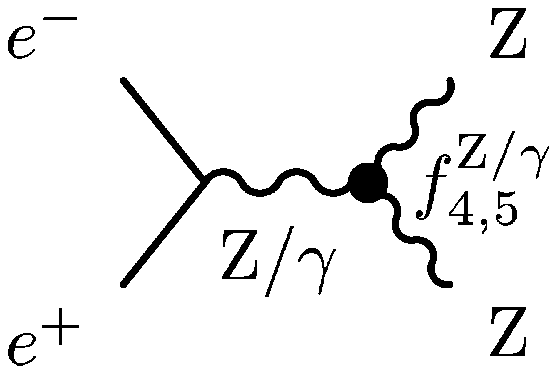} 
\hspace{0.15\textwidth}
\includegraphics*[height=0.15\textwidth]{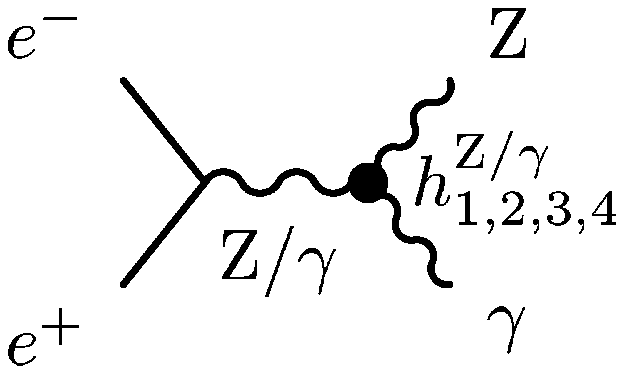}
\caption{Couplings of three neutral bosons: Anomalous vertices.}
\label{fig:tnc}
\end{figure}

Couplings of three neutral bosons do not exist in the \SM. By imposing
only Lorentz and \Uem\ invariance, and for final states with equal
bosons Bose symmetry, one ends up with possible anomalous vertices
shown in Fig.~\ref{fig:tnc}. The corresponding
Lagrangians~\cite{hagiwara,gounaris} describing these anomalous
vertices are
\begin{displaymath}
\begin{array}{rcl}
\mathcal{L}_{NP}^{VZZ} = \frac{e}{m_Z^2}
\Bigl[
&-& {\red   f_4^V} (\partial_\mu V^{\mu\beta}) Z_\alpha(\partial^\alpha Z_\beta)
 +  {\green f_5^V} (\partial^\sigma V_{\sigma\mu}) \tilde{Z}^{\mu\beta}Z_\beta
\Bigr] \\
\mathcal{L}_{NP}^{VZ\gamma} = \frac{e}{m_Z^2}
\Bigl[
&-& {\blue  h_1^V} (\partial^\sigma V_{\sigma\mu}) Z_\beta F^{\mu\beta}
 -  {\cyan  h_3^V} (\partial_\sigma V^{\sigma\rho})Z^\alpha \tilde{F}_{\rho\alpha}\\
&-& 
\frac{{\magenta h_2^V}}{m_Z^2}[\partial_\alpha \partial_\beta (\Box + m_V^2) V_\mu]
Z^\alpha F^{\mu\beta}
+
\frac{{\orange h_4^V}}{2m_Z^2}[(\Box+m_V^2)\partial^\sigma V^{\rho\alpha}] Z_{\sigma}\tilde{F}_{\rho\alpha}
\Bigr],
\end{array}
\end{displaymath}
with $\tilde{V}_{\mu\nu} = \nicefrac{1}{2} \;
\epsilon_{\mu\nu\rho\sigma} V^{\rho\sigma}$ and $V = {\rm Z},
\gamma$. The Lagrangians are of higher order than those for the gauge
couplings of the W boson, so that one would expect either to detect
deviations more easily with the W boson couplings or the scale of New
Physics (which is artificially set to $m_Z$ in the above formulae) to
be close. The couplings {\red $f_4^V$}, {\blue $h_1^V$} and {\magenta
$h_2^V$} are \CP\ violating, whereas the couplings {\green $f_5^V$},
{\cyan $h_3^V$} and {\orange $h_4^V$} conserve \CP. One interesting
option for the future, which has not been followed yet, is to relate
the couplings through {\magenta \SU\ symmetry}~\cite{alcaraz-2}. This
relates the couplings from the Z$\gamma$ and from the ZZ final state
in the following way: ${\green f_5^{\rm V}} = {\cyan h_3^{\rm V}} \tan
\theta_{\rm W}$ and ${\red f_4^{\rm V}} = {\blue h_1^{\rm V}} \tan
\theta_{\rm W}$.

\begin{figure}
\centerline{\includegraphics*[width=0.65\textwidth]{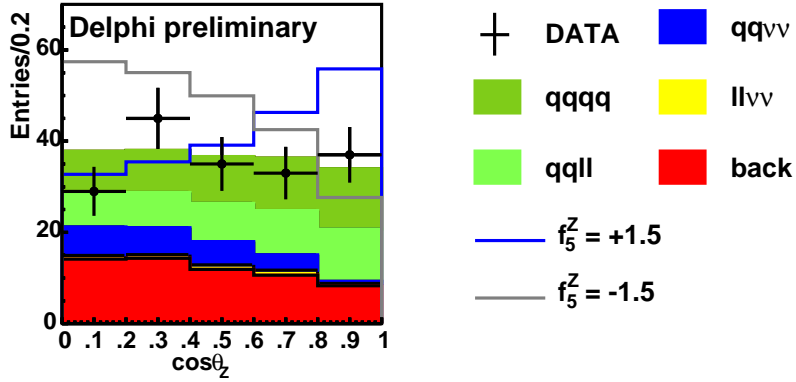}}
\caption{ZZ production angle measured by \DELPHI\ compared to the \SM\
prediction and $f_5^{\rm Z} = \pm 1.5$.}
\label{fig:delphi_ngc_zz}
\end{figure}

The measurement of the $f$ couplings proceeds by selecting events from
all visible ZZ final states and then reweighting the distributions for
different values of the anomalous couplings $f_{4,5}^{{\rm
Z},\gamma}$. In the presence of anomalous couplings, the total
cross-section, the production angle of the Z boson and the average
polarization of the Z bosons would change. In
Fig.~\ref{fig:delphi_ngc_zz} the distribution of the Z boson
production angle $\cos \theta_{\rm Z}$ as predicted by the \SM\ and
for $f_5^{\rm Z} = \pm 1.5$ is compared to the data, as they have been
measured by the \DELPHI\ experiment.

Since in all \LEP\ data no evidence for the presence of anomalous $f$
couplings has been found, limits at the 95\% confidence level are
set. These limits are derived either one-dimensional by fixing all
other couplings to zero, or two-dimensional by fitting couplings with
the same \CP\ behavior at the same time. The one-dimensional limits
are\cite{lep-gc}:
\begin{displaymath}
-0.17 < f_4^\gamma < 0.19\hspace{1cm}
-0.31 < f_4^Z      < 0.28\hspace{1cm}
-0.36 < f_5^\gamma < 0.40\hspace{1cm}
-0.36 < f_5^Z      < 0.39
\end{displaymath}

\begin{figure}[b]
\includegraphics*[width=0.99\textwidth]{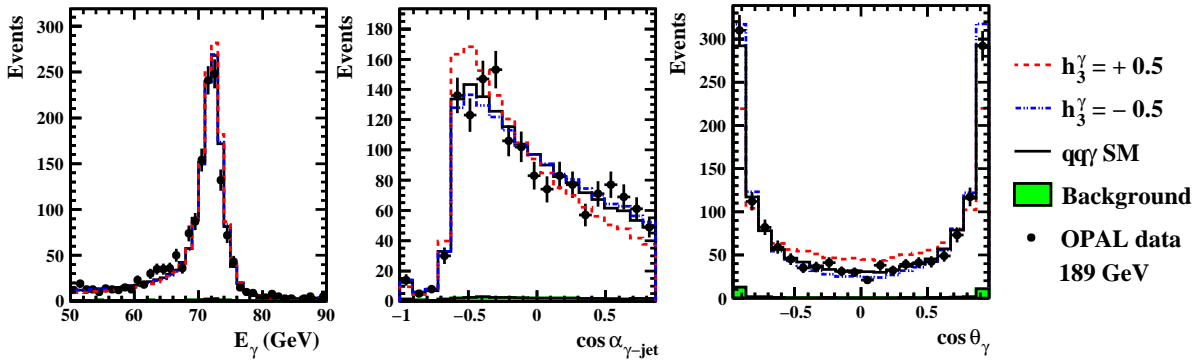}
\caption{Distributions of $E_\gamma$, $\cos \alpha_{\gamma-{\rm jet}}$
and $\cos\theta_\gamma$ for the \qqgamma\ final state. Predictions
from the \SM\ and for $h_3^{\gamma} = \pm 0.5$ are compared to the
data.}
\label{fig:opal_qqg_shape}
\end{figure}

For the $h$-couplings, events of the reactions $\eeZgto \qqgamma$ and
$\eeZgto \nngamma$ are selected. The photon energy $E_\gamma$, the
angle $\cos\alpha_{\gamma-{\rm jet}}$ between the photon and the
nearest jet, and the photon production angle $\cos\theta_{\gamma}$ are
sensitive to the anomalous couplings. In
Fig.~\ref{fig:opal_qqg_shape}, distributions of these variables from
the \OPAL\ experiment are shown, for the \SM\ prediction and for
$h_3^{\gamma} = \pm 0.5$. No evidence for anomalous $h$ couplings has
been found, and one- and two-dimensional limits are derived. The
one-dimensional limits are\cite{lep-gc}:
\begin{displaymath}
\begin{array}{llll}
-0.056 < h_1^\gamma < 0.055 &
-0.045 < h_2^\gamma < 0.025 &
-0.130 < h_1^Z      < 0.130 &
-0.078 < h_2^Z      < 0.071   \\
-0.049 < h_3^\gamma < 0.008 &
-0.002 < h_4^\gamma < 0.034 &
-0.200 < h_3^Z      < 0.070 &
-0.050 < h_4^Z      < 0.120
\end{array}
\end{displaymath}

\section{Quartic boson self couplings}

Starting from \Uem\ gauge invariance and requiring a custodial
SU(2)$_c$ symmetry, genuine quartic couplings (i.\,e. quartic
couplings that are not introduced to counteract the trilinear gauge
couplings to achieve \SU\ symmetry) arise through the
Lagrangians\cite{belanger,stirling}

\begin{displaymath}
\begin{array}{ll}
\mathcal{L}_0 = - \frac{e^2}{16} \frac{{\red   a_0^{W,Z}}}{{\orange \Lambda}^2} F^{\mu\nu} F_{\mu\nu} \vec{W}^\alpha \vec{W}_\alpha & WW\gamma\gamma, ZZ\gamma\gamma\\
\mathcal{L}_c = - \frac{e^2}{16} \frac{{\green a_c^{W,Z}}}{{\orange \Lambda}^2} F^{\mu\alpha} F_{\mu\beta} \vec{W}^\beta \vec{W}_\alpha & WW\gamma\gamma, ZZ\gamma\gamma\\
\mathcal{L}_n = - \frac{e^2}{16} \frac{{\blue  a_n}}{{\orange \Lambda}^2} \vec{W}_{\mu\alpha} \cdot (\vec{W}_{\nu} \times \vec{W}^{\alpha}) F^{\mu\nu} & WWZ\gamma
\end{array}
\end{displaymath}

\begin{figure}
\includegraphics*[width=0.2\textwidth]{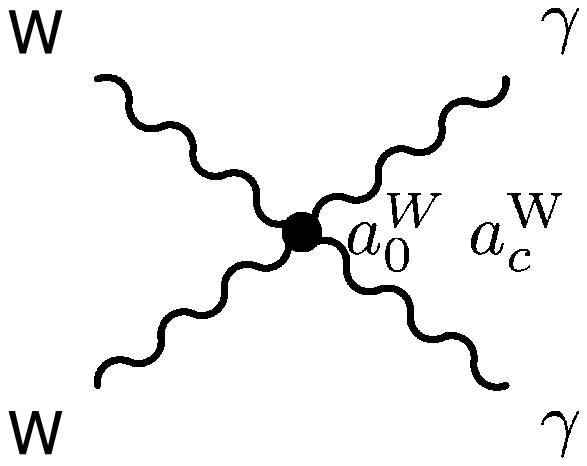}\hspace{3cm}
\includegraphics*[width=0.2\textwidth]{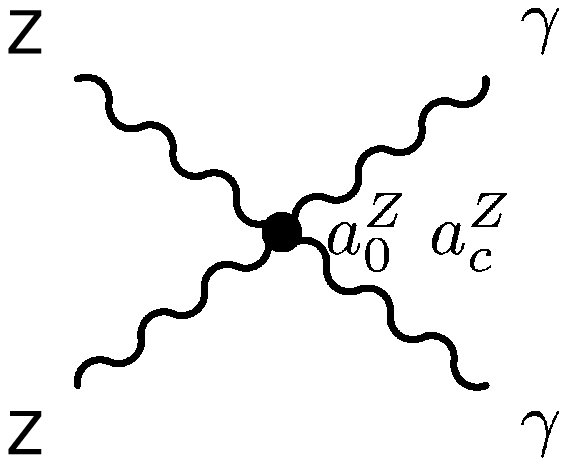}\hspace{3cm}
\includegraphics*[width=0.2\textwidth]{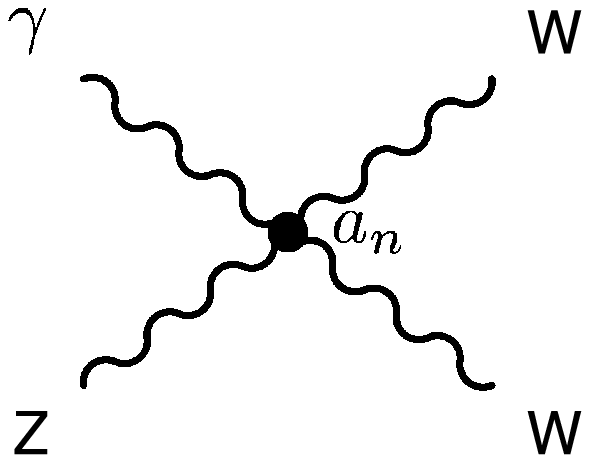}
\caption{Anomalous contributions to quartic gauge couplings.}
\label{fig:quartic-anom}
\end{figure}

The couplings $a_0$ and $a_c$ conserve \CP, the coupling $a_n$
violates \CP. Figure~\ref{fig:quartic-anom} shows the relationship
between the vertices and the anomalous couplings. In principle, the
couplings of the W can be different from the couplings of the Z, hence
the different superscripts.

These couplings are accessible either through boson fusion with two
bosons in the final state or through the production of three gauge
bosons. The fusion processes become important only at Linear Collider
energies and are negligible at \LEP. Recent results\cite{l3-wwg} from
\LD\ for the process $\eeto W^+ W^- \gamma$, which would dominate a
possible \LEP\ combination for $a^{\rm W}_0$, $a^{\rm W}_c$ and $a_n$,
allow to set the following limits at 95\% \CL:
\begin{displaymath}
\begin{array}{ccc}
-0.02 < {\red   a^{\rm W}_0} / {\orange \Lambda}^2 \cdot \mbox{GeV}^2 < 0.02 &
-0.05 < {\green a^{\rm W}_c} / {\orange \Lambda}^2 \cdot \mbox{GeV}^2 < 0.03 &
-0.14 < {\blue  a_n} / {\orange \Lambda}^2 \cdot \mbox{GeV}^2 < 0.13
\end{array}
\end{displaymath}
The energy of the least energetic photon in the process $\eeto
Z\gamma\gamma$ is especially sensitive to the presence of anomalous
quartic couplings and used as a test distribution. Since no evidence
for such couplings is found, limits are set at 95\% \CL\ by L3
as\cite{l3-zgg}:
\begin{displaymath}
\begin{array}{cc}
-0.02 < {\red   a^{\rm Z}_0} / {\orange \Lambda}^2 \cdot \mbox{GeV}^2 < 0.03 &
-0.07 < {\green a^{\rm Z}_c} / {\orange \Lambda}^2 \cdot \mbox{GeV}^2 < 0.05
\end{array}
\end{displaymath}

\section*{Acknowledgments}

I would like to thank my colleagues from \LD\ and the convenors of the
\LEP\ four-fermion and gauge-coupling working groups for their support
and many fruitful discussions.


\end{document}